\documentclass[a4paper,10pt]{article}

\usepackage{amsmath}
\usepackage{graphicx}
\usepackage{amsfonts}
\usepackage{amssymb}
\usepackage{color}
\usepackage{comment}
\usepackage{ulem}
\usepackage{tikz}

\newcommand{\bc}{\begin{center}}
\newcommand{\ec}{\end{center}}
\def\ba#1{\begin{array}{#1}\displaystyle}
\newcommand{\ea}{\end{array}}

\newcommand{\beq}{\begin{equation}}
\newcommand{\eeq}{\end{equation}}
\newcommand{\beqa}{\begin{eqnarray}}
\newcommand{\eeqa}{\end{eqnarray}}

\newcommand{\bi}{\begin{itemize}}
\newcommand{\ei}{\end{itemize}}

\def\lt#1{\left#1}
\def\rt#1{\right#1}
\def\t#1{\tilde{#1}}

\def\b#1{\bar{#1}}
\def\frc#1#2{\frac{#1}{#2}}

\newcommand{\R}{{\mathbb{R}}}

\newcommand{\ep}{\epsilon}

\begin{document}

\begin{center}
{\Large {\bf Non-Equilibrium Conformal Field Theories\\ \vskip 0.2 truecm with Impurities}}
\vskip 1.0 truecm

{\large D. Bernard${}^{\clubsuit}$, B. Doyon${}^{\spadesuit}$, J. Viti${}^{\diamondsuit}$}

\vspace{0.5cm}
{\small ${}^{\clubsuit}$ Laboratoire de Physique Th\'eorique de l'ENS, CNRS $\&$ Ecole Normale Sup\'erieure de Paris, France.}\\
{\small ${}^{\spadesuit}$ Department of Mathematics, King's College London, London, United Kingdom.}\\
{\small ${}^{\diamondsuit}$  Max Planck Institute For Complex Systems, Dresden, Germany.}\\
\end{center}

\vspace{1.0cm} 
\noindent{\bf Abstract}\\
We present a construction of non-equilibrium steady states within conformal field theory. These states sustain energy flows between two quantum systems, initially prepared at different temperatures, whose dynamical properties are represented by two, possibly different, conformal field theories connected through an impurity. This construction relies on a real time formulation of conformal
defect dynamics based on a field scattering picture parallelizing -- but yet different from -- the Euclidean formulation. We present the basic characteristics of this formulation and give an algebraic construction of the real time scattering maps that we illustrate in the case of $SU(2)$-based conformal field theories.
\vskip 1.5 truecm

\section{Introduction}

Conformal Field Theory (CFT) \cite{BPZ,DiF-Mat-Sen} has in particular been developed to characterize one-dimensional gapless quantum systems at equilibrium. Driven by substantial experimental advancements, see \cite{Revue} for a comprehensive survey, the focus is nowadays on non-equilibrium quantum phenomena and dynamics. For instance, it has become possible to accurately measure the current flowing between two leads driven out-of-equilibrium through voltage or temperature gradient \cite{Mesure,Pierre} and to  implement transport experiments with cold-atomic gases~\cite{KSH14}.
Although a lot of theoretical progress has been achieved in the realm of non-equilibrium physics over the past years, especially for low dimensional systems, see e.g.~\cite{GLSS06, AGMT09, GGM10, MS13, GT14} and \cite{Zwanzig} for a review, an unifying framework extending the methods of statistical physics out-of-equilibrium  is still lacking and any simple model of a genuine system far from equilibrium  could guide in deciphering properties and principles underlying these phenomena.

In this note, our aim is to construct non-equilibrium steady states (NESS) within the formalism of conformal field theory, as workable and hopefully useful examples of far from equilibrium steady states. We will specialize to conformal field theory, a method \cite{Ruelle} formulated to study  out-of-equilibrium quantum systems. We refer the reader to \cite{BD12,BD13} for a more detailed discussion of such a  setup. It amounts to imagine preparing at different temperatures two quantum systems, called the left and the right parts and which are assumed to be described by two conformal field theories (or more precisely, whose low energy sectors are assumed to be describable by two conformal field theories), and gluing them together through a contact point, called the impurity. Once the contact has been established, the energy flows from the hotter to the colder part of the system, and a non-equilibrium steady state is eventually reached after a transient period. 
We will present a characterization of these steady states exploiting tools of conformal field theory.  In \cite{BD12,BD13}, an explicit construction was given when the left and right halves were identical and the contact such that the whole system is homogeneous after the gluing so that the energy in this case can flows without reflexion. Here, we extend such a construction to the general situation where the left and right halves are associated to different conformal field theories and the contact, being non-homogeneous, induces transmission and reflexion of the energy flow.

The paper is organized as follows. The characterization of the non-equilibrium steady states given in Section~\ref{Sec:NESS} requires defining real time dynamics for conformal field theories with defects. We formulate this definition in Section~\ref{Sec:dynamics}. The effects of the defects are specified by so-called dual defect maps coding for the field scatterings on the impurity which we characterize in Section~\ref{Sec:Charac}. Section~\ref{Sec:Fermion} is an illustration of this framework in the case of free fermions. The principle for an algebraic construction of the real time defect dynamics is presented, and illustrated in the case of $SU(2)$, in Section~\ref{Sec:Moduli}. 

\section{Real time dynamics and dual defect maps}\label{Sec:dynamics}

Constructing non-equilibrium steady states within the aforementioned framework requires dealing with infinitely large subsystems and real time dynamics, as otherwise the steady regime would not be attained. Hence we must consider conformal field theories on non-compact intervals connected through impurities and the real time dynamics associated to those defects. Defects in conformal field theories have of course already been considered in the physics literature, including \cite{PZ01,F+07,QRW07,BB08,DKR11}. However, these studies dealt with Euclidean conformal field theories and we found that we had to adapt the construction to real time conformal dynamics.

\subsection{Kinematics}

We consider a system modeled by an impurity placed at the origin $x=0$ of the real line $\R$, and generically different CFTs on the left ($x<0$) and the right ($x>0$).
We use throughout the upper-indices $*^{l,r}$ to refer to quantities on the left and right sides, respectively, of the impurity. For simplicity we restrict ourselves to cases in which the chiral and anti-chiral operator algebras of the CFTs are isomorphic.
We denote these operator algebras by ${V}^{l,r}$ (for chiral fields) and $\bar V^{l,r}$ (for anti-chiral fields), respectively. They are isomorphic to the left/right vertex operator algebras ${\cal V}^{l,r}$, with  $V^l\simeq \bar V^l\simeq {\cal V}^l$ and $V^r\simeq \bar V^r\simeq {\cal V}^r$. As usual in the context of
vertex operator algebras, these are specified by infinite-dimensional vector spaces of operators $\widehat{\varphi}^{l,r}_a\in {\cal V}^{l,r}$, and by structure constants
$(C^{l,r})_{aa'}^{a''}$ coding for the Operator Product Expansions (OPEs)\footnote{We shall reserve the ``hat" notation to fields related to the abstract vertex operator algebras,
and the ``hat-less" one to fields within the CFTs quantized on the real line.} \cite{BPZ,DiF-Mat-Sen} :
\beq\label{ope}
	\widehat{\varphi}^{l,r}_a(z)\, \widehat{\varphi}^{l,r}_{a'}(w) =\sum_{a''} (C^{l,r})_{aa'}^{a''}\, (z-w)^{h_{a''}^{l,r}-h_{a'}^{l,r}-h_a^{l,r}}\, \widehat{\varphi}^{l,r}_{a''}(w). 
\eeq
Operator algebras are modules, generically not irreducible, for the Virasoro algebras generated by $\widehat L_n^{l,r}$,
\[
	[\widehat L_m^{l,r},\widehat L_n^{l,r}] = (m-n)\widehat L_{m+n}^{l,r} + \frc{c^{l,r}}{12} (m^3-m)\delta_{m+n,0},
\]
where $c^{l,r}$ are the left/right central charges and $h_a^{l,r}$ are $\widehat L_0^{l,r}$-eigenvalues (conformal dimensions).  The series on the right-hand side of \eqref{ope} may be interpreted as an infinity of products between $\widehat{\varphi}^{l,r}_a$ and $\widehat{\varphi}^{l,r}_{a'}$. The identification of the chiral and anti-chiral operator algebras $V^{l,r}$ and $\bar V^{l,r}$ with the vertex operator algebra ${\cal V}^{l,r}$ has to be chosen to have clear properties under the hermitian conjugation corresponding to quantization on the line, rather than that on the circle naturally associated to the normalization in \eqref{ope}. Given $\varphi^{l,r}\in {\cal V}^{l,r}$,  this normalization is implemented by\footnote{This may be obtained by mapping the sphere to the cylinder of radius $R$, using the conformal $x\to z=e^{\frac{2\pi i}{R}x}$, and then sending the radius of the cylinder to infinity. The fields are then transported via the conformal rules: $\phi(x)=\lim_{R\to \infty} \big(\frac{2\pi}{R}\big)^h e^{2\pi i x h/R}\, \widehat \phi(z=e^{2\pi i x/R})$ for the chiral field, and $\bar \phi(x) =\lim_{R\to \infty}\big(\frac{2\pi}{R}\big)^{\bar h}e^{-2\pi i x {\bar h}/R}\, \widehat {\bar \phi}(\bar z=e^{-2\pi i x/R})$ for the anti-chiral fields, with $h$ and $\bar h$ the conformal dimension. The OPEs (\ref{ope-Phi}) may be viewed as a consequence of the fact that $[(\frac{R}{2\pi}) (e^{i\pi(x-y)/R}-e^{-i\pi(x-y)/R})]=[i(x-y)]$ as $R\to\infty$. } 
$\widehat{\varphi}^{l,r} \mapsto \phi^{l,r} = e^{-\frc{i\pi}2 \hat L_0^{l,r}} \widehat{\varphi}^{l,r}\in V^{l,r}$ and $\widehat{\varphi}^{l,r} \mapsto \bar\phi^{l,r} = e^{+\frc{i\pi}2\hat{\b L}_0^{l,r}} \hat{\varphi}^{l,r}\in \bar V^{l,r}$. The normalized CFT fields $\phi^{l,r}$ and $\bar \phi^{l,r}$ then satisfy the OPEs ($x,y\in\R$):
 \begin{eqnarray}\label{ope-Phi}
  \phi^{l,r}_a(x)\phi^{l,r}_{a'}(y) &=& \sum_{a''} (C^{l,r})_{aa'}^{a''}\, [i(x-y)]^{h_{a''}^{l,r}-h_{a'}^{l,r}-h_a^{l,r}}\, \phi^{l,r}_{a''}(y),\\
 \bar \phi^{l,r}_a(x)\bar \phi^{l,r}_{a'}(y) &=& \sum_{a''} (C^{l,r})_{aa'}^{a''}\, [-i(x-y)]^{h_{a''}^{l,r}-h_{a'}^{l,r}-h_a^{l,r}}\,\bar \phi^{l,r}_{a''}(y).
  \end{eqnarray}
In particular, for the stress tensor components $T$ and $\bar T$ one has (by definition of a CFT~\cite{BPZ,DiF-Mat-Sen}),
\begin{eqnarray}\label{ope-T}
T^{l,r}(x)\, \phi(y) &=& \sum_n \, [i(x-y)]^{-n-2}\, (L^{l,r}_n\phi)(y),\\
\bar T^{l,r}(x)\, \bar \phi(y) &=& \sum_n\, [-i(x-y)]^{-n-2}\, (\bar L^{l,r}_n\bar \phi)(y),
\end{eqnarray}
with $L_n^{l,r}$ and $\bar L_n^{l,r}$ the Virasoro generators acting on the chiral and anti-chiral fields, respectively. Recall that the space of fields is in one-to-one correspondence with the vertex operator algebras. 
We will use the notations $T(x) := T^{l,r}(x)$ and $\b T(x) := \b T^{l,r}(x)$ for $x\lessgtr 0$.

\subsection{Dynamics}

The dynamics can be seen as a flow $t\mapsto U_t,\;t\in\R$, on the operator algebra of the product of the two CFTs. We denote the
time evolution of fields by $U_t[\varphi(x)] = \varphi(x,t)$, and we take $t>0$ in the following. We define it in the simplest possible way: it is ballistic except for possible scattering on the impurity.

\begin{figure}[t]
\centering
\begin{tikzpicture}[scale=1.0]
\draw[thick,->](-1,-1)--(-0.5,-0.5);
\draw[thick](-0.5,-0.5)--(0,0);
\draw[thick,->](1,-1)--(0.5,-0.5);
\draw[thick](0.5,-0.5)--(0,0);
\draw[thick](-0.5+1,-0.5+1)--(0+1,0+1);
\draw[thick,->](-1+1,-1+1)--(-0.5+1,-0.5+1);
\draw[thick,->](1-1,-1+1)--(0.5-1,-0.5+1);
\draw[thick](0.5-1,-0.5+1)--(0-1,0+1);
\draw[yellow, fill=yellow!50] (-0.1,-1) rectangle (0.1,1);
\draw[->](-2,-1)--(2,-1);
\draw(2,-1) node[above] {$x$};
\draw[->](-2,-1)--(-2,-0.5);
\draw(-2,-0.5) node[left] {$t$};
\draw(-1,-1) node[below]{$\bar{V}^l$};
\draw(1,-1) node[below]{$V^r$};
\draw(-1,1) node[above]{$V^l$};
\draw(1,1) node[above]{$\bar{V}^r$};
\draw(-0.1,0) node[left]{$\Theta$};
\end{tikzpicture}
\caption{The defect OPE is assumed to allow for a scattering picture
of local fields belonging to the vector spaces $\bar{V}^l$ and $V^r$. This is a schematic picture,
we will be more precise in the text.}
\label{figevol}
\end{figure}
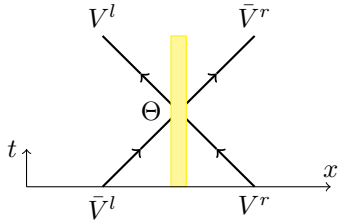


By locality of the time evolution and conformality on the left and on the right, away from the impurity the dynamics is chiral.
That is $\phi(x,t)=\phi(x-t)$ and $\bar \phi(x,t)=\bar \phi(x+t)$ (for right-movers and left-movers resp.), as long as the
translated fields ``do not hit the impurity'', namely for all $t$ such that $x$ and $x\mp t$ (for right-movers and left-movers resp.) have the same sign. 
For values of $t$ such that the translated chiral field would cross the impurity, the dynamics is modified. The modification of the dynamics can in principle be deduced using OPEs between local bulk fields and the defect line.
Such OPEs can be written as series involving defect fields, \textit{i.e.} local fields positioned on the defect line.
Given a defect line, many bulk-field configurations are expected to lead to the same series involving defect fields, and this gives rise to equivalences between bulk-field configurations near defect lines.
Our main assumption is that such equivalences allow for a \textit{field scattering picture}:
it is possible to identify right-moving fields just on the right of the impurity with combinations of left-moving fields on the right side of the impurity and right-moving fields on the left side  (see Figure \ref{figevol});  a similar identification holds for left-moving fields just on the left of the impurity.
Since, upon chiral time evolution $U_t$ with positive time, one never finds left/right-moving fields just on the right/left of the impurity, no other identification is necessary in order to define the evolution map. The argument is extended to the simultaneous presence of right-moving and left-moving fields on the right and left, respectively, and the scattering on the impurity is then encapsulated into a dual defect map\footnote{We call it ``dual defect map'' in order to stress its connection with defect maps, or operators, already introduced in the literature on {\it Euclidean} conformal field theory. Another name could have been ``impurity scattering map".}
\beqa  \label{defTheta}
\Theta: \bar V^l \otimes V^r \to V^l \otimes \bar V^r.
\eeqa
With explicit indices this reads
\begin{equation}\label{indexTheta}
\Theta\, [\bar\phi^l_a\, \phi^r_b]=\sum_{cd}\Theta_{ab}^{cd}\ \phi^l_c\, \bar \phi^r_d.
\end{equation}
According to this assumption, the full time evolution can be obtained from the map $\Theta$, so that it is sufficient to characterize $\Theta$ without the necessity of analyzing defect OPEs.
We expect our setup to have large validity.

The complete time evolution is defined by combining the chiral evolution and the defect map, see Figure \ref{figevol}.
For clarity, let us repeat the argument. Away from the impurity the behavior is chiral, so that
\begin{eqnarray}\label{chiral_left}
\phi^l(x,t) &:=&\phi^l(x-t),\hskip 1.0 truecm \mathrm{for}\ x<0,\ \mathrm{all}\ t,\\
\bar \phi^l(x,t)&:=&\bar \phi^l(x+t),\hskip 1.0 truecm \mathrm{for}\ x<0,\ x+t<0,\nonumber
\end{eqnarray}
and
\begin{eqnarray}\label{chiral_right}
\phi^r(x,t) &:=& \phi(x-t),\hskip 1.0 truecm \mathrm{for}\ x>0,\  x-t>0, \nonumber\\
\bar \phi^r(x,t)&:=&\bar \phi(x+t),\hskip 1.0 truecm \mathrm{for}\ x>0,\  \mathrm{all}\ t. 
\end{eqnarray}
The scattering on the impurity is represented by the dual defect map, so that
\begin{equation} \label{ThetaAB}
\bar \phi^l_a(-x,t)\, \phi^r_b(x,t) := \sum_{cd} \Theta_{ab}^{cd}\ \phi^l_c(x-t)\, \bar \phi^r_d(t-x) = {\cal E}_{t-x}\big(\Theta[\b\phi_a^l \phi_b^r]\big),\quad  \mathrm{for}\  t>x>0,
\end{equation}
where for convenience we introduced the notation
\beqa \label{defEt}
	{\cal E}_x[\phi^{l,r}] := \phi^{l,r}(-x),\quad {\cal E}_x[\bar \phi^{l,r}] := \bar\phi^{l,r}(x),
\eeqa
for $\phi^{l,r}$ (resp. $\bar \phi^{l,r}$) any chiral (resp. anti-chiral) fields, and we extend it on products naturally.
For our purposes, it is enough to consider products of fields localized symmetrically at opposite positions: by chiral evolution,
they are indeed going to hit the impurity at the same instance, and one can certainly specialize in eq.\eqref{ThetaAB} either $\bar \phi^l_a$ or $\phi^r_b$ to the identity field. 

By definition, the time evolution defines a flow on the operator algebra of the product CFT, so we extend the map $U_t$ on product of fields at various positions by tensoring eqs.\eqref{chiral_left}, \eqref{chiral_right} and \eqref{ThetaAB}. As we shall see below, demanding this flow to be an automorphism of the operator algebra imposes constraints on the dual defect map.

Finally, an important property of the dynamics, usually understood as the expression of conformal invariance, is the local conservation of energy at the impurity\footnote{Here, we assume that the impurity possesses no relevant degree of freedom, and hence it carries no energy.}. The energy density is $e(x,t):=T(x,t)+\bar T(x,t)$. Chirality ensures local energy conservation away from the impurity, i.e. $\partial_t e(x,t)=\partial_x p(x,t)$ with $p(x,t):=T(x,t)-\bar T(x,t)$ the momentum density. Conservation of the energy at the impurity location constrains dual defect maps, as we explain below.

\section{Characterizing the dual defect map}\label{Sec:Charac}

\subsection{Boundary conditions and intertwining relation}
Conservation of the energy requires that the momentum operator $p(x,t) := T(x,t)-\bar T(x,t)$ is continuous at the origin at any time,
therefore we can set $p(0,t):=p(0^-,t) = p(0^+,t)$. This condition, which is equivalent to conformal invariance, reads
\begin{equation} \label{conserv}
 T(0^-,t)+\bar T(0^+,t)= T(0^+,t)+\bar T(0^-,t).
 \end{equation}
Deciphering its consequences requires a little care because the operator algebra is defined on the line at fixed time, while the above equation involves  stress energy tensor components at fixed position but different times.
We shall prove that the local continuity equation (\ref{conserv}) is equivalent to
\beq\label{cond}
	\Theta \circ ( \b L_n^l+ L_n^r) = (  L_n^l + \b L_n^r ) \circ \Theta.
\eeq
Recall that $\Theta$ and the Virasoro generators $L_n^{l,r}$ and $\bar L_n^{l,r}$ are all defined as maps acting on operators, so that the relation \eqref{cond} must also be thought as a map between operator spaces. The equation (\ref{cond}) shows that $\Theta$ intertwines between the Virasoro representations defined on $\bar V^l\otimes V^r$ and  $V^l\otimes \bar V^r$. The total central charge is $c^l+c^r$ in both tensor products, and by construction these representations are isomorphic, with a trivial isomorphism being $\b\phi_a^l \phi_{a'}^r \mapsto \phi_a^l\b\phi_{a'}^r$. Demanding $\Theta$ to be invertible, a consequence of the fact that it describes a unitary evolution, ensures that it is a generically nontrivial isomorphism.
The proof of eq.(\ref{cond}) goes in two steps: we first derive a global version of the local continuity condition (\ref{conserv}) and deduce eq.(\ref{cond}) as a compatibility condition
between such a global continuity equation and operator product expansions.
 
{\it A global version of the continuity equation.~}
Let us first show that the continuity equation (\ref{conserv}) together with the chiral evolution away from the impurity is equivalent to (here, we do not need the definition of the dual defect map):
\beq\label{cont}
T(x,t)+\bar T(-x,t)= T(x-t)+\bar T(-x+t)
\eeq
 
{\it Proof:} First taking $x=0^+$ (and $t>0$) in eq.(\ref{cont}) we get $T(0^+,t)+\bar T(0^-,t)=T(-t)+\bar T(t)$. By chirality (away from the origin), we have $T(0^-,t)=T(-t)$ and $\bar T(0^+,t)=\bar T(t)$. Thus we get $T(0^+,t)+\bar T(0^-,t)= T(0^-,t)+\bar T(0^+,t)$ which is the continuity equation $p(0^-,t) = p(0^+,t)$.\\
Second, let us look at $T(x,t)+\bar T(-x,t)$ with $x>0$, $t=x+s>x$. Assuming that the continuity equation $p(0^-,t) = p(0^+,t)$ holds, we have:
\begin{eqnarray*}
T(x,t)+\bar T(-x,t) &=& T(0^+,s)+\bar T(0^-,s)\quad ~~~~~~~~~~ \mathrm{(by\ chirality)}\\
&=& T(0^-,s)+\bar T(0^+,s) \quad ~~~~~~~~~~ \mathrm{(by\ momentum\ continuity)}\\
&=& T(x-t)+\bar T(-x+t) \quad ~~~~~~~ \mathrm {(by\ chirality)}
\end{eqnarray*}
Finally eq.(\ref{cont}) is clearly true by chirality for $x>0$ and $t<x$ and for all $t$ if $x<0$.
\hfill $\square$
\medskip

{\it Time evolution and operator algebra automorphism.~} 
Let us now prove that eq.(\ref{cond}) follows from the assumption that the time evolution preserves the operator algebra together with the momentum continuity condition (here we only assume the time evolution preserves OPEs with the stress-tensor). 

{\it Proof:} We test the global continuity equation (\ref{cont}) against field localized at points $y$ and $-y$ on each side of the impurity. It is enough to test it on products of chiral fields $\bar \phi^l(-y)\phi^r(y)$, with $y>0$ by convention. We start
with the product $(T(x)+\bar T(-x))\bar\phi^l(-y)\phi^r(y)$. Since, by hypothesis, the time evolution preserves OPEs with the stress energy tensor components,
we can rewrite the global continuity equation (\ref{cont}) as
\[
U_t[(T(x)+\bar T(-x))\cdot \bar\phi^l(-y)\phi^r(y)]=(T(x-t)+\bar T(-x+t))\cdot \bar\phi^l(-y,t)\phi^r(y,t).
\]
We can now look at the OPEs of both sides of this equation, recalling how the stress energy tensor acts on fields.
Let us extract the coefficients of $[i(x-y)]^{-n-2}$, which we denote by $(l.h.s.)_n$ and $(r.h.s.)_n$.\\
On the l.h.s. (recall that we assume $x,y>0$) we have $(l.h.s)_n= U_t[ (L^r_n +  \bar L^l_n) \bar\phi^l(-y)\phi^r(y)]$.
For $t>y>0$ this involves the map $\Theta$ because the chiral fields hit the impurity (after a time $y$). Once they have crossed the impurity (via the action of the map $\Theta$) their evolution is chiral. Thus
\[ (l.h.s.)_n= ({\cal E}_{t-y}\circ \Theta\circ(L^r_n + \bar L^l_n)) [\bar\phi^l\,\phi^r]. \]
These are fields localized at $t-y$ or $-y+t$ depending on their chirality.\\
On the r.h.s. we first time-evolve the product $\bar\phi^l(-y)\phi^r(y)$ and then do the OPE with the stress tensors. For $t>y>0$ we have 
$\bar\phi(-y,t)\phi(y,t)= ({\cal E}_{t-y}\circ \Theta)\, [ \bar\phi^l\,\phi^r]$.
Again this gives fields localized at $t-y$ or $-y+t$ depending on their chirality. Doing the OPE we get
\[ (r.h.s.)_n= ({\cal E}_{t-y}\circ(L^r_n + \bar L^l_n)\circ \Theta)\, [ \bar\phi^l\,\phi^r]. \]
Comparing $(l.h.s.)_n$ and $(r.h.s.)_n$ we obtain eq.(\ref{cond}). 
\hfill $\square$
\subsection{Operator algebra automorphism}\label{Sect:auto}

The time evolution must be an operator algebra automorphism. This property was used in the previous paragraph to derive eq.\eqref{cond}, but it implies further constraints on the dual defect map.
A first feature is that $\Theta$ must preserve the identity field
\beq\label{thid}
	\Theta({\bf 1}\otimes {\bf 1}) = {\bf 1}\otimes {\bf 1}.
\eeq
Along with eq.(\ref{cond}) this actually implies the momentum continuity equation.
Indeed, choosing $n=-2$ in eq.\eqref{cond} and applying it on ${\bf 1}\otimes {\bf 1}$ gives $\Theta[\b T^l + T^r] = T^l + \b T^r$. Applying then ${\cal E}_t$ and exploiting
the time-evolution equations (\ref{chiral_left}, \ref{chiral_right}) and \eqref{ThetaAB} we recover the momentum conservation \eqref{conserv}. 

Furthermore demanding $\Theta$ be an automorphism for the product
of the left/right chiral algebra ${\cal V}^l\times{\cal V}^r$ ensures that the time evolution $U_t$ preserves the
operator algebra of the product CFT.
Fields in ${\cal V}^l\times{\cal V}^r$ are linear combinations
of products $\widehat \varphi_{ab}:=\widehat \varphi^l_a\,\widehat \varphi^r_b$ with conformal
weight $h_{(ab)}=h_a+h_b$, and the structure constants
of their OPEs  are products of the left/right structure constants.
Checking that the time evolution specified by such a map $\Theta$
is an automorphism of the operator algebra on the line is straightforward.
One has to consider the time evolution of product of nearby fields,
symmetrically localized and crossing the impurity at large enough time.
That is, one has to look at the products
$\bar \phi^l_a(-x,t)\phi^r_b(x,t)\cdot \bar \phi^l_{a'}(-y,t)\phi^r_{b'}(y,t)$ for $t>x,y>0$,
and compare the two OPEs obtained either by time evolving the series resulting for the OPEs of the initial time fields
or by expanding the OPEs of the time evolved fields.
The proof is then simply a matter of writing these OPEs and
noticing that
the factors $i$ and $(-i)$ in the OPEs of the chiral and anti-chiral fields conspire
with the reflexion $x\to -x$ on the impurity.

A third characteristic is that demanding the time evolution to be an operator automorphism ensures that the map $\Theta$ is uniquely defined by its restrictions on the left anti-chiral fields and on the right chiral fields separately (when embedding $\bar V^l$ and $V^r$ into $\bar V^l\otimes V^r$ by tensoring them with the identity operator, i.e. $\bar V^l\hookrightarrow \bar V^l\otimes {\bf 1}$ and $V^r \hookrightarrow{\bf 1}\otimes V^r$). This is a direct consequence of the facts that the limit $\lim_{\ep\to0}\b\phi^l_a(-x-\ep)\phi^r_{a'}(x)$ is well defined and equal to $\b\phi^l_a(-x)\phi^r_{a'}(x)$ and that the resulting equality may be time-evolved. In other words, consistency between OPEs and time evolution enforces the dynamics to be fully defined once it is specified on the left/right chiral or anti-chiral fields.
\subsection{The pure reflection case}

In the pure reflection case the defect disconnects the left/right CFTs. Let $\Theta_0$ be a dual defect map for pure reflection.
Since the two halves are separated, it factorizes into two maps $\Theta^l_0$ and $\Theta^r_0$ acting respectively
from $\bar V^l$ to $V^l$ and from $V^r$ to $\bar V^r$. That is, $\Theta_0 = \Theta_0^l\otimes \Theta_0^r$ and
$\Theta^r_0[T^r]=\bar T^r$ and $\Theta^l_0[\bar T^l]=T^l$. This is equivalent to the statement that no energy
flows through the origin, $T(0^+,t)=\bar T(0^+,t)$ and $\bar T(0^-,t)=T(0^-,t)$, as it should be.

Further, eq.(\ref{cond}) implies $\Theta_0^r L^r_n= \bar L_n^r\Theta_0^r$ and $\Theta_0^l \bar L^l_n= L_n^l\Theta_0^l$.
Let us assume that the spaces of chiral operators are decomposable as sum of irreducible Virasoro modules
(or of irreducible modules of some extended algebra), 
that is $V^{l,r}=\oplus_j R_j^{l,r}$ with $R_j^{l,r}$ irreducible representations of some (extended) conformal symmetry algebra.
Similarly $\bar V^{l,r}=\oplus_j \bar R_j^{l,r}$ with $\bar R_j^{l,r}$ isomorphic to $R_j^{l,r}$ (recall that we assume
for simplicity that $\bar V^{l,r}$ is isomorphic to $V^{l,r}$). By Schur's lemma, $\Theta_0^{l,r}$ is then a linear 
combination of the equivariant maps identifying $\b R_j^{l,r}$ with  $R_j^{l,r}$ 
\begin{equation} \label{Theta-reflex}
\Theta_0^l=\sum_j \zeta_j^l\, \mathbb{I}_{\bar R_j^l\to R_j^l},\quad
\Theta_0^r=\sum_j \zeta_j^r\, \mathbb{I}_{R_j^r\to \b R_j^r}.
\end{equation} 
As a consequence, there is no mixing between the primary operators when reflected at the boundary.
This translates into the following evolution equations for the primary fields which we denote $\Phi^l_j$
\begin{equation} \label{reflex-phi}
\bar\Phi^l_j(-x,t) = \zeta^l_j\, \Phi^l_j(x-t)
,\quad \Phi^r_j(x,t) = \zeta^r_j\, \b \Phi^r_j(-x+t)
,\quad \mathrm{for}\ x>0,\ t>|x|.
\end{equation}
Consistency with OPEs then yields stringent constraints on the parameters
$\zeta^{l,r}_j$. Let $(C^{l,r})_{jk}^m$ be the structure constants relating the primary fields
$\Phi^{l,r}_j$, $\Phi^{l,r}_k$ and $\Phi^{l,r}_m$ (these are the constants involved in the fusion rules).
The requirement that $\Theta$ be an automorphism of the operator algebra imposes that $\zeta^{l,r}_j\zeta^{l,r}_k=\zeta^{l,r}_m$ for all triplets $j,k,m$ such that $(C^{l,r})_{jk}^m\neq0$. Since the identity operator is reflected trivially, we have $\zeta^{l,r}_0=1$. Let us denote $\Phi^\dag_{j} = \Phi_{\t j}$ using the hermitian structure on the line, then $\zeta_{\t j}^{l,r} = (\zeta_{j}^{l,r})^*$. Since $(C^{l,r})_{j\t j}^0\neq0$, we then find $|\zeta^{l,r}_j|=1$. Thus for self-conjugated representation $\zeta^{l,r}_j=\pm1$, and in general the $\zeta^{l,r}_j$ are numbers of modulus one. These phases are then recursively constrained by the fusion rules.

\subsection{Connection with Euclidean CFT conformal defects}

We now mention elements of the relation between this {\it real time} formulation and the more usual
{\it Euclidean} formulation in which a defect \cite{PZ01,F+07,QRW07,BB08,DKR11} between two conformal
field theories $CFT^1$ and $CFT^2$ is specified through a map $D$ from, and into, the CFT's bulk operator spaces,
i.e. $D: {\cal H}_1\to {\cal H}_2$ where ${\cal H}_j$ refer to the bulk operator Hilbert spaces of the two CFTs.
In radial quantization the condition for conformal invariance is then written as 
\begin{equation} \label{E-defect}
({\cal L}_n^2-\b {\cal L}_{-n}^2)\, D = D\, ({\cal L}_n^1-\b {\cal L}_{-n}^1).
\end{equation}
where $\mathcal{L}^{1,2}_n$ are the Virasoro generators acting respectively on $\mathcal{H}^{1,2}$.

The connection between these two formulations is through a crossing relation (as suggested by diagrammatic pictures). 
Recall that $\Theta: \bar V^l \otimes V^r \to V^l \otimes \bar V^r $ satisfies the intertwining relation (\ref{cond}).
Let ${}^{\dagger^{l\times r}}$ denote the hermitian conjugation acting on both $V^l$ and $V^r$,
but not on $\bar V^l$ and $\bar V^r$. 
Let $D_{l;r}$ be the map defined by conjugation
\[ D_{l;r}:= \mathrm{const.}\,\Theta^{\dagger^{l\times r}}: V^{l\, *}\otimes \bar V^l \to V^{r\, *}\otimes \bar V^r.\]
If the Virasoro representations on $V^l$ and $V^r$ are unitary in the sense
that $(L_n^{l,r})^{\dagger^{l\times r}}=L_{-n}^{l,r}$, it immediately follows that eq.~(\ref{cond}) for $\Theta$ is equivalent to: 
\[ D_{l;r}\, (L^l_n-\bar L_{-n}^l) = (L^r_n-\bar L_{-n}^r) \, D_{l;r}.\]
Hence, on each irreducible components of the left/right CFTs, $D_{l;r}$ is proportional to a standard Euclidean conformal defect,
up to exchanging the chiral representations with their contragradiants; reciprocally, $\Theta= \mathrm{const.}\, D_{l;r}^{\dagger^{l\times r}}$. Notice that the map $(L_n,c)\to (-L_{-n},-c)$ is an automorphism of the Virasoro algebra so that $(L^{l,r}_n-\bar L_{-n}^{l,r})$ define representations of the Virasoro algebra on $V^{l,r}\otimes\bar V^{l,r}$ but of zero central charge.

The normalizing coefficients are fixed by imposing  compatibility with OPEs and fusion rules and/or with modular invariance. The pure reflexion case indicates that the maps $D$ and $\Theta$ differ by normalizations and these relative normalizations can furthermore be different in different irreducible sectors. Indeed, in this case, $\Theta_0=\sum_j\zeta_j\, \mathbb{I}_{R_j\to \hat R_j}$ with $\zeta_j$ of modulus one and constrained by consistency with the OPEs, as discussed above, whereas the Euclidean defect operators are labeled by the space of primary fields and read $D^{(k)}=\sum_j \frac{S_{kj}}{\sqrt{S_{0j}}} \mathbb{I}_{R_j\otimes\bar R_j\to \mathbb{C} }$ with $S_{kj}$ the modular matrix transformation, as discussed in \cite{Cardy_bdry}. We do not have yet a simple explanation of these differences to offer. 

\section{Non-Equilibrium Steady States}\label{Sec:NESS}

Non-equilibrium CFT steady states are defined as in \cite{BD12,BD13} following the approach of \cite{Ruelle, PilletEtAl}.
The data are two dynamical evolutions:
the one when the two left/right parts are decoupled and
that occurring when they are coupled by a non-trivial dual defect map.
We denote the first by $U_t^0$ and the second by $U_t$.
One starts with a state $\omega_0$, stationary with respect to $U_t^0$,
and we shall take $\omega_0$ to be the tensor product of the two left/right Gibbs states at respective
temperatures $T_l$ and $T_r$. One then imagines time evolving
this state with the coupled left/right dynamics $U_t$ for a long period of time.
The non-equilibrium steady state is obtained by sending to infinity this time interval. Hence, 
provided that  the limit exists, one defines
\[ \omega_{\rm ness}(\Phi)=\lim_{t\to\infty}\omega_0(U_t[\Phi]),\]
for any product of operators $\Phi$. By construction $\omega_{\rm ness}$ is, if it exists, $U_t$-stationary. Since $\omega_0$ is $U_t^0$-stationary, we have $\omega_0({U_t^0}^{-1}[\hat \Phi])=\omega_0[\hat \Phi]$
for any $\hat \Phi$, and hence $\omega_0(U_t[\Phi])=\omega_0({U_t^0}^{-1}U_t[\Phi])$ by choosing
$\hat \Phi$ to be $U_t[\Phi]$. Taking the large time limit
and defining the $S$-matrix by $S[\Phi]:= \lim_{t\to\infty} {U_t^0}^{-1}\, U_t\, [\Phi]$
(assuming the limit exists) allows us to express $\omega_{\rm ness}$  as \cite{Ruelle}
\beqa \label{ness0}
\omega_{\rm ness}( \Phi) = \omega_0( S[\Phi] )
\eeqa
The $S$-matrix, and hence $\omega_{\rm ness}$, is completely, and explicitly, defined by the two dual defect maps $\Theta_0$ and $\Theta$ respectively associated to the dynamical evolution $U_t^0$ and $U_t$. 
On chiral and anti-chiral fields, the $S$-matrix acts as follows
\beqa 
S[ \phi(x) ] &=& \lt\{\ba{ll}
\phi(x),& \hskip 0.8 truecm(x<0) \\
({\cal E}_{-x}\circ  \Theta_0^{-1}\Theta)[ \mathbb{I}\otimes \phi ], & \hskip 0.8 truecm (x>0)
\ea\rt. \\
S[ \bar\phi(x) ] &=& \lt\{\ba{ll}
({\cal E}_{x}\circ  \Theta_0^{-1}\Theta)[ \bar\phi\otimes \mathbb{I} ],& \hskip 1.0 truecm (x<0)\\
\bar\phi(x), & \hskip 1.0 truecm (x>0)
\ea\rt. \label{Sbar}
\eeqa

The mean energy current is $J_E:= \omega_{\rm ness}[T(x,t)-\bar T(x,t)]$ because $p(x,t):=T(x,t)-\bar T(x,t)$ is the momentum operator. Conservation of energy ensures that it is $x$-independent, and it is time independent by stationarity of $\omega_{\rm ness}$. 
As it will become explicit below, the mean energy current is non vanishing. This implies that the state $\omega_{\rm ness}$  breaks time reversal invariance, because the momentum operator is odd under time reversal, and thus that it is out of equilibrium. The steady state also sustains a positive entropy production. This can be argued for following standard thermodynamical arguments. Indeed according to thermodynamics, the entropy production per unit of time is $\sigma:={T_r}^{-1}(\frac{dE^r}{dt})+{T_l}^{-1}(\frac{dE^l}{dt})$ with $E^{l,r}$ the mean energies of the left/right sub-systems (reservoirs) and $T_{l,r}$ their respective temperatures. By energy conservation, $\frac{dE^r}{dt}=-\frac{dE^l}{dt}=J_E$. Hence the entropy change per unit of time is $\sigma=({T_r}^{-1}-{T_l}^{-1})\, J_E$, which is positive since $J_E\propto (T_l^2-T_r^2)$, see eq.(\ref{currentK}). It turns out that this expression also follows from a more accurate definition of the entropy production for quantum steady states of the type we are studying, based on the notion of relative entropy and studied in the context of $C^*$-algebra descriptions \cite{Ruelle2,Pillet1,Jak1,Jak2} (see also the review \cite{bensaad}).

\section{The free fermions case}\label{Sec:Fermion}

Let us make a pause with a simple case: free fermions with central charge $c=1/2$.
Let $\psi(x)$ and $\bar \psi(x)$ be the two chiral components of a Majorana fermion $(\psi=\psi^\dag)$. 
Their OPEs on the line reads
\begin{align} 
\label{OPE_ff1}
\psi(x)\psi(y)&\simeq (-i)\Big[ \frac{1}{x-y}- 2(x-y) T(y)+\cdots \Big], \\
\label{OPE_ff2}
\bar \psi(x)\bar \psi(y)&\simeq (+i)\Big[ \frac{1}{x-y}- 2(x-y) \bar T(y)+\cdots \Big] .
\end{align}
They satisfy anti-commutation relations, $\{\psi(x),\psi(y)\}=2\pi\delta(x-y)$ and $\{\bar{\psi}(x),\bar{\psi}(y)\}=2\pi\delta(x-y)$, with operator ordering defined by the usual $i0^+$ time ordering prescription. The stress tensor components are $T= -\frac{i}{2} (\partial \psi)\psi$ and $\bar T= \frac{i}{2} (\partial \psi)\psi$.

The impurity rotates the fermions so that the dynamics is, by definition
\begin{eqnarray} 
\label{dyn_free1}
 & ~ & \lt\{\ba{ll} \psi^r(x,t) := \cos\alpha\cdot \psi^l(x-t) + \sin\alpha\cdot \bar \psi^r(-x+t),\quad & (x>0,\ t>|x|),\\
\bar \psi^l(x,t) := \cos\alpha\cdot \bar \psi^r(x+t) - \sin\alpha\cdot \psi^l(-x-t),\quad &(x<0,\ t>|x|).
\ea\rt. 
\end{eqnarray}
It preserves the OPEs and hence the anti-commutation relations.
The map $\Theta$ can be read from eq.(\ref{dyn_free1}):
$\Theta[\psi^r]=\cos\alpha\, \psi^l + \sin\alpha\, \bar \psi^r$ and
$\Theta[\bar \psi^l]=\cos\alpha\, \bar\psi^r-\sin\alpha\, \psi^l$.
The pure transmitting case corresponds to $\alpha=0$ and the pure reflexion to $\alpha=\pi/2$ and this defines $\Theta_0$.
 
The energy current $J_E$ is $x$-independent and we can compute it by locating the
fields at  either sides of the origin, so that $J_E=\omega_0[S[T(x)-\bar{T}(x)]]$ with $x>0$ or $x<0$ equivalently.
Let us choose $x>0$. The $S$-matrix acts trivially on $\bar T(x)$ and non-trivially on $T(x)$. For $x>0$,
we have $S[\bar{T}^r(x)]=\bar{T}^r(x)$ and
\begin{equation*}
S[T^r(x)]=\cos^2\alpha~\bar{T}^l(-x)+\sin^2\alpha~T^r(x)-
\frac{i}{2}\cos\alpha\sin\alpha\,[\partial_x\bar{\psi}^l(-x)\psi^r(x)-\partial_x\psi^r(x)\bar{\psi}^l(-x)]
\end{equation*}
The state $\omega_0$ is Gaussian when expressed in terms of the
fermionic modes, and therefore only operators with an even number
of fermions at the same point have non-zero expectation value.
Using $\omega_0(T^{l,r}(x))=\frac{c \pi}{12}T_{l,r}^2$ with $c=1/2$ for free Majorana fermions, we conclude that
\begin{equation}
 J_E=\frac{\pi\,\cos^2\alpha}{24}\, \left(T_l^{2}-T_r^{2}\right).
\end{equation}
In the pure transmitting case $\alpha=0$, we recover $J_E=\frac{\pi\,c}{12}\, \left(T_l^{2}-T_r^{2}\right)$ with central charge $c=1/2$ \cite{BD12,BD13}.

As an illustration of the previous statements, we can verify that \eqref{cont} is satisfied
by the physical stress tensor in the free fermion case. Recall that $\Theta$, and hence the time evolution,
is an automorphism of the operator algebra so that
\begin{align*}
&T(x,t)=-\frac{i}{2}[\cos\alpha~\partial_x\psi(x-t)-\sin\alpha~\partial_x\bar{\psi}(-x+t)][\cos\alpha~\psi(x-t)+\sin\alpha~\bar{\psi}(-x+t)]\\
&\bar{T}(-x,t)=+\frac{i}{2}[\cos\alpha~\partial_x\bar{\psi}(-x+t)+\sin\alpha~\partial_x{\psi}(x-t)][\cos\alpha~\bar{\psi}(-x+t)-
\sin\alpha~\psi(x-t)];
\end{align*}
therefore  $T(x,t)+\bar{T}(-x,t)=T(x-t)+\bar{T}(-x+t)$.

In the free fermion case, the space of chiral fields has the structure of Fock spaces. Let us introduce fermionic operators
$b^{l,r}_s$, $s\in \mathbb{Z}+1/2$, with $\{b_s^{l,r},b_{s'}^{l,r}\}=\delta_{s+s';0}$, and similarly for $\bar b^{l,r}_s$.
The fermionic Fock space is obtained by applying the operators $b^{l,r}_{-s}$ to the identity ${\bf 1}$ with $b^{l,r}_s{\bf 1}=0$ for $s>0$. By definition $\psi^{l,r}=b^{l,r}_{-1/2}{\bf 1}$ and similarly for $\bar \psi^{l,r}$. The Virasoro algebra acts on those Fock
spaces via $L^{l,r}_n=\frac{1}{2}\sum_{m\in\mathbb{Z}} m :b_{n-m+1/2}^{l/r}b_{m-1/2}^{l,r}:$ and similarly
for $\bar L^{l,r}_n$. This action is such that $(L^{l,r}_{-1})^n\psi = n!\, b^{l,r}_{-n-1/2}{\bf 1}$
Since $L_{-1}$ acts as a $x$-derivative, consistency with (\ref{dyn_free1}) yields
\begin{equation}
\Theta\left[\begin{array}{c}
             b_{n+1/2}^r\\
           \bar b_{n+1/2}^l\end{array}
\right]=\left[
\begin{array}{cc}
 \cos \alpha & \sin\alpha \\
 -\sin\alpha & 	\cos \alpha
\end{array}\right]\left[\begin{array}{c}
             b_{n+1/2}^l\\
            \bar b_{n+1/2}^r\end{array}
\right]\Theta.
\end{equation}
Hence, the map $\Theta$ acts on all the fermionic modes as a Bogoliubov transformation, and it is a standard exercise to check that it intertwines the Virasoro representations.

\section{An algebraic construction of dual defect maps}\label{Sec:Moduli}

\subsection{The moduli space}

The fermonic case described above has a simple interpretation:
the tensor product $\bar V^l\otimes V^r$ supports a representation of the Virasoro algebra of central
charge $c=1$ and is isomorphic to the Gaussian free field representation; it thus has a global $U(1)$ symmetry
commuting with the diagonal Virasoro action but not with each of the Virasoro actions on $\bar V^l$ or $V^r$; 
the map $\Theta$, which acts as $O(2)$-rotation on the fermions, is the action induced by this symmetry. 

It is clear that such procedure is more general: 
any automorphism of the product vertex operator algebra ${\cal V}^l\times{\cal V}^r$ commuting with
the diagonal Virasoro action but acting non trivially on each of the components ${\cal V}^l$ or ${\cal V}^r$ yields
a non trivial map $\Theta$. The moduli space of such maps is 
\begin{equation} \label{moduli}
\frac{ \rm{Aut}\, [{\cal V}^l \times {\cal V}^r]}{\rm{Aut}\, [{\cal V}^l]\times \rm{Aut}\, [{\cal V}^r]}
\end{equation}
Thus, to construct a conformally invariant dynamics with an impurity we have to look for conformally equivariant automorphisms of ${\cal V}^l \times {\cal V}^r$ mixing the left and right vertex operator algebras. Given such automorphism the dual defect map is defined by identifying ${\cal V}^l \times {\cal V}^r$ with $\bar V^l\otimes V^r$ for the incoming fields and with $V^l\otimes\bar V^r$ for the outgoing fields. The statement proved in Section \ref{Sect:auto} ensures that the defect dynamics is then consistent.

A simple way to realize this construction consists in considering two CFTs, the left and the right, which are not individually invariant under a group $G$ but whose product is. For instance, one may choose the left CFT to be a WZW model \cite{DiF-Mat-Sen} on a group $H\subset G$ and the right CFT to be the parafermionic coset theory $G/H$ for some group $G$ and adjust the levels of each of those CFTs such that the product CFT is isomorphic the WZW model on $G$ and hence is $G$ invariant.

\subsection{The $SU(2)_{k}$ case}

A simple, but actually quite generic, example can be given. Choose the left CFT to be $U(1)_k$ at a specified radius $R_k$ and the right CFT to be the $\mathbb{Z}_k$-parafermionic theory \cite{DiF-Mat-Sen}. Choose $R_k$ such that $U(1)_k\times\mathbb{Z}_k\simeq SU(2)_k$, that is, such that the tensor product of the left and right CFTs is isomorphic to the $SU(2)$-WZW model at level $k$. By construction the product $\mathcal{V}^l\times \mathcal{V}^r$ then possess a global $SU(2)$ symmetry commuting with the total Virasoro generators. We can choose $\Theta$ to be the automorphism induced by any element of that $SU(2)$. Equivalently, picking a group element $g\in SU(2)$, the map $\Theta$ is that induced by this global $SU(2)$-rotation. However, since rotations corresponding to the distinguished $U(1)$ subgroup are symmetries of $\mathcal{V}^l$ and $\mathcal{V}^r$ independently, the moduli space of this construction is $SU(2)/U(1)$, which is isomorphic to the two-sphere.

Let us first restrict ourselves to the $k=2$ case, i.e. to the case
$SU(2)_{k=2}\simeq U(1)\times \mathbb{Z}_2$ with central charge $c=3/2=1+1/2$. 
Thus we consider a $c=1$ CFT (at radius $R=1$) with $U(1)$ symmetry on the left and a $c=1/2$ free Majorana
fermion CFT on the right. Let $\widehat J$ be the $U(1)$ current and $\widehat \psi$ be the free fermion. 
They generate $\mathcal{V}^l\times \mathcal{V}^r$ by definition. The $SU(2)_{k=2}$-currents are represented as
\[ \widehat J^\pm=\widehat \psi\, \widehat V_\pm,\quad \widehat J^0 = \widehat J,\]
where we used the notation $\widehat V_\pm:=e^{\pm i\hat X}$ for the bosonic vertex operators with
$\widehat J:=i\partial \hat X$. Primary fields are $\widehat \varphi_j^m$, with $j=0,\frac{1}{2},1$, with conformal weight
$h_j=\frac{j(j+1)}{k+2}$, i.e. $h_0=0$, $h_{\frac{1}{2}}=\frac{3}{16}$ and $h_1=\frac{1}{2}$. We only need the $j=1$ primary. Let $\widehat \varphi^0:=\widehat \varphi_{j=1}^{m=0}$ and $\widehat \varphi^\pm:= \widehat \varphi_{j=1}^{m=\pm1}$. They may be represented as
\[ \widehat \varphi^0 = \eta\, \widehat \psi,\quad \widehat \varphi^\pm= \pm\eta\, \widehat V_\pm,\]
with $\eta$, $\eta^2=1$, commuting with $\widehat \psi$ and $\widehat J$ but anti-commuting with $\widehat V_\pm$.
Under a global $SU(2)$ transformation,
both operator multiplets $(\widehat J^0, \widehat J^\pm)$ and $(\widehat \varphi^0,\widehat \varphi^\pm)$ transform similarly
because they both take values in the adjoint representation. Hence, under a global $SU(2)$ transformation,
\begin{eqnarray}\label{rotateJ}
\widehat J^0 \mapsto s\, \widehat J^0 + \frac{1}{\sqrt{2}} \big( \bar r\, \widehat J^+ + r\, \widehat  J^- \big),\quad
\widehat \varphi^0 \mapsto  s\,\widehat \varphi^0 + \frac{1}{\sqrt{2}} \big( \bar r\, \widehat \varphi^+ + r\, \widehat \varphi^- \big),
\end{eqnarray}
with $s^2+r\bar r=1$. Transformations of the other components of the current  and $j=1$ primary multiplets are similarly
defined by $SU(2)$ adjoint action (for a given group element). Of course one can check that these transformations
preserve all OPEs. We shall use the parametrization
\[ s:= \sin\alpha,\quad r:= e^{i\beta}\, \cos\alpha.\]

The map $\Theta$ is induced from these transformations by the (now usual) embeddings of the operator algebras
$\mathcal{V}^l$ (resp. $\mathcal{V}^r$) onto the set of chiral and anti-chiral operators $V^l$ and $\bar V^l$ (resp. $V^r$ and $\bar V^r$). It specifies a conformally invariant dynamics in presence of a defect connecting the $c=1$ left CFT and the $c=1/2$ CFT. For the $U(1)$ current and the fermion the induced time evolution reads (with the identification $\widehat \psi \eta \hookrightarrow \psi$)
\begin{eqnarray} \label{eq:jsu2-time}
\bar J(x,t) &=& s\, J(-x-t) + \frac{1}{\sqrt{2}}\big( \bar r\, V_+(-x-t) + r \,V_-(-x-t)\big) \bar \psi(x+t),\hskip 0.5 truecm \mathrm{for}\ x<0,\ t>|x|,\\ \label{eq:psu2-time}
\psi(x,t) &=& s\, \bar\psi(-x+t) + \frac{1}{\sqrt{2}}\,\eta\big( \bar r\, V_+(-x-t) - r\, V_-(-x-t)\big), \hskip 1.55 truecm \mathrm{for}\ x>0,\ t>|x|.
\end{eqnarray}
The left/right stress tensors are
$T^l=\frac{1}{2} J^2$ and $T^r=-\frac{i}{2}(\partial\psi)\psi$, and it is  easy to check that the global conservation law
(\ref{cont}) is fulfilled thanks to (\ref{eq:jsu2-time},\ref{eq:psu2-time}). 

One could continue working directly with the $SU(2)$ formalism but we can also use a short cut by
fermionizing the left CFT \cite{SenFerm}. Let us define
\begin{equation*} 
\chi_1(x)=\frac{1}{\sqrt{2}}\big( e^{i\beta}V_+(x)+ e^{-i\beta} V_-(x)\big) ,\quad
\chi_2(x)=\frac{1}{\sqrt{2}}\,\eta \big( e^{i\beta}V_+(x)- e^{-i\beta} V_-(x)\big),
\end{equation*}
These operators are real $(\chi_j^\dagger=\chi_j$) fermions with OPEs $\chi_j(x)\chi_k(y)\simeq [i(x-y)]^{-1}\, \delta_{jk}+\cdots$. The equations (\ref{eq:jsu2-time},\ref{eq:psu2-time}) are then compatible with a pure reflexion for $\chi_1$ and a rotation between $\psi$ and $\chi_2$. 
Thus, $\chi_1$ decouples and the energy flow is only transported by the modes $\chi_2$ and $\psi$.
We are back to previous computation. In particular, the mean energy current is
\begin{equation*}
 J_E = \frac{\pi}{12} \big(\frac{r\bar r}{2} \big)\big(T_l^2 - T_r^2\big).
\end{equation*}

The advantages of using the group theoretical approach is that it yields all information on the dynamics and guarantees consistency. It can easily be generalized to any level $k$, as we now briefly illustrate by computing the mean energy current. In that case one has a $U(1)$ CFT on the left, with central charge $c^l=1$, and a $\mathbb{Z}_k$ parafermionic CFT on the right, with central charge $c^r=\frac{2(k-1)}{k+2}$. In terms of these data, the $SU(2)_k$ currents are represented by:
\[ \widehat J^0= \sqrt{{k}/{2}}\, \widehat J,\quad \widehat J^\pm= \sqrt{k/2}\, \widehat \Psi_{\pm1}\, e^{\pm i \sqrt{\frac{2}{k}} \hat X},\]
with $\widehat J=i\partial \hat X$, the $U(1)$ current, and $\widehat \Psi_{\pm 1}$ the parafermionic field \cite{DiF-Mat-Sen}. The $SU(2)_k$ total stress tensor can be written either as the sum of the $U(1)$ and $\mathbb{Z}_k$ stress-tensors or using the Sugawara construction\cite{DiF-Mat-Sen}. Hence,
\[ 
\widehat T_{su(2)}=\frac{1}{k+2}\big( \hat J^0 \hat J^0 +  \hat J^+ \hat J^-+ \hat J^- \hat J^+\big) = \widehat T_{u(1)}+\widehat T_{\mathbb{Z}_k}.
\]
where $\widehat T_{u(1)}=\frac{1}{2} \widehat J^2$ and
$\widehat T_{\mathbb{Z}_k}$ are the $U(1)$ and $\mathbb{Z}_k$ stress tensors, respectively. Let us now implement a global
$SU(2)$ transformation as in (\ref{rotateJ}) so that: $\widehat J^0 \mapsto  s\, \widehat J^0 + \frac{1}{\sqrt{2}} (\bar r\, \widehat J^+ + r\, \widehat  J^-)$ with $s^2+r\bar r=1$. This yields a transformation of the $U(1)$ stress tensor
\[ \widehat T_{u(1)} \mapsto \big(s^2+\frac{r\bar r}{k}\big)\, \widehat T_{u(1)} + \big(\frac{k+2}{2k}\big)(r\bar r)\, \widehat T_{\mathbb{Z}_k}+\cdots,\]
where the dots refer to operators with non-zero $U(1)$ charges. 
By construction, this immediately gives the way the $S$-matrix acts on the left stress-tensor.
Namely, for $x<0$,  $S[T^l(x)]=T^l(x)$ and 
\[ S[ \bar T^l(x)] = \big(s^2+\frac{r\bar r}{k}\big)\, \bar T^l(x) + \big(\frac{k+2}{2k}\big)(r\bar r)\, T^r(-x) + \cdots,\]
where again the dots refer to charged operators. Recall that the mean energy current
is $J_E=\omega_0[S[T(x)-\bar{T}(x)]]$. Since charged operators have vanishing expectations with respect to
$\omega_0$, and using $\omega_0(T^{l,r}(x))=\frac{\pi}{12}c^{l,r}T^2_{l,r}$, we get
\begin{equation} \label{currentK}
J_E=\frac{\pi}{12}\big(\frac{k-1}{k}\big)(r\bar r)\, \big(T^2_l-T^2_r\big).
\end{equation}
It is remarkable that the mean energy current is always proportional to the difference of the temperatures squared.

\section{Conclusion}

We have described a framework for dealing with non-equilibrium CFTs with impurities.
It is based on a real time formulation of CFTs with
defects\footnote{One may also wonder about possible connection between the dual defect maps and
solutions of the Yang-Baxter equation. However, we do not have yet a simple answer to this question.}.
The algebraic construction of the dual defect maps that we illustrated in the $SU(2)_{k}$ case can be generalized to other groups.
This will most probably be done elsewhere \cite{BDVxx},
including the evaluation of the full counting statistics large deviation functions.
The generalization of this construction to impurities with intrinsic degrees of freedom is worth considering,
as is the connexion with Euclidean conformal defects.

The relevance, or irrelevance, of the CFT description to out-of-equilibrium physical situations is a question worth asking
because dangerous irrelevant operators may play a role and modify the low energy description \cite{Giama92}.
However, to our knowledge, there is no consensus in the existing literature on whether
transport in 1D quantum systems is predominantly ballistic or diffusive at low-enough temperature.
Some time ago, ref.\cite{Cast95} claimed that energy transport has a  ballistic nature 
in any integrable systems. This suggestion has later received further support by the work \cite{Pros11}, where an 
exact bound on the so-called Drude weight was derived for integrable systems driven by Lindblad
operators. Numerical confirmations, although limited to  XXZ spin chains,
have also been found in  \cite{Moore12, DVMR14} and the very recent \cite{MCJ14}.
Since a CFT is integrable,
this would imply a non-negligible ballistic component even for non-integrable microscopic 1D systems if their low-temperature
regimes were described by non-equilibrium CFTs. However, ref.\cite{Aff05} claimed that diffusion is universal,
even for integrable systems. Surprising, non-zero Drude weights in non-integrable systems were also found in \cite{Moore12}, implying
a component of ballistic transport for these systems as well.
Finally, the $T^2$ dependence of the energy current, that
is a sign of ballistic transport, has been experimentally verified \cite{Pierre} in mesoscopic systems of
sizes smaller than their coherence lengths.

These intriguing facts indeed suggest -- or give support to the claim -- that a ballistic channel for energy transport exists at low temperatures and that CFTs describe these out of equilibrium regimes.
However, the scaling limit is never exact but only asymptotic.
Thus, there may -- and actually there should as we would like to argue --
exist a large relaxation time $T_{\rm relax}$ (or equivalently, a large characteristic
scale $\ell_{\rm relax}= v_FT_{\rm relax}$, with $v_F$ the Fermi velocity, which we may identify with the coherence length)
depending on microscopic data above which transport becomes diffusive.
That is, we  expect transport to be essentially ballistic at low temperatures
and at times large but smaller than $T_{\rm relax}$, and to become diffusive at times larger than
$T_{\rm relax}$ for non-integrable models. Recall that the size of the samples used in the experiments \cite{Pierre}
are smaller than the decoherence length.
The ballistic behaviors seen in \cite{Moore12} for non-integrable systems is probably due to the fact that the simulation
times were long enough to reach quasi-stationary states but smaller than the relaxation time
$T_{\rm relax}$ above which the systems would have behaved diffusively.
We believe that the ballistic quasi-stationary states
should be describable by non-equilibrium conformal field theories.
\bigskip 

{\bf Acknowledgements:} This work was supported in part by the ANR contracts ANR-2010-BLANC-0414 and ANR-14-CE25-0003-01.
DB and BD are glad to respectively thank Jean-Bernard Zuber and Gerard Watts for discussions. JV would like to thank
Jerome Dubail and Jean-Marie Stephan for helpful comments and Andrea de Luca for collaboration at the early stages of the project.


\begin{thebibliography}{}

\bibitem{AGMT09} K. Andrieux, P. Gaspard, T. Monnai, S. Tasaki, ``The fluctuation theorem for currents in open quantum systems", New J. Phys. 11 (2009), 043014

\bibitem{PilletEtAl} W.H. Aschbacher, C.-A. Pillet, ``Non-equilibrium steady states of the XY chain", J. Stat. Phys. 112, 1153 (2003)

\bibitem{BB08} C. Bachas, I. Brunner, ``Fusion of conformal interfaces", JHEP02 (2008) 085.

\bibitem{BPZ} A.A. Belavin, A.M. Polyakov, A.B. Zamolodchikov, ``Infinite conformal symmetry in two-dimensional quantum field theory", Nucl. Phys.B241, 333 (1984).

\bibitem{bensaad} R. Ben S‰ad, C.-A. Pillet, ``A geometric approach to the Landauer-BŸttiker formula'', J. Math. Phys. 55, 075202 (2014)

\bibitem{BD12} D. Bernard, B. Doyon, ``Energy flow in non-equilibrium conformal field theory.", J. Phys. A 45, 362001 (2012).

\bibitem{BD13} D. Bernard, B. Doyon, ``Non-equilibrium steady-states in conformal field theory", Ann. H. Poincare, online first, arXiv:1302.3125.

\bibitem{BDVxx} D. Bernard, B. Doyon, J. Viti, ``Non-equilibrium CFT (II)", in preparation.

\bibitem{Revue} Y.M. Blanter, M. Buttiker, ``Shot noise in mesoscopic conductors", Phys. Rep. 336 (2000) 1.

\bibitem{Mesure} J. Bylander, T. Duty, and P. Delsing, ``Current measuring by real time counting of single electrons", Nature 434, 361 (2005).

\bibitem{Cast95} H. Castela, X. Zotos, P. Prelovsek, ``Integrability and Ideal Conductance at Finite Temperatures", Phys. Rev. Lett. {\bf 74} (1995) 972.

\bibitem{Cardy_bdry} J. Cardy, ``Conformal Invariance and Surface Critical Behavior", Nucl. Phys. B240 [FS12], 514-532, 1984.

\bibitem{DKR11} A. Davydov, L. Kong,  I. Runkel, ``Invertible defects and isomorphisms of rational CFTs", Adv. Math. Theor. Phys. 15 (2011), 43.

\bibitem{DiF-Mat-Sen} Ph. Di Francesco, P. Mathieu, D. Senechal, ``Conformal Field Theory", Springer (New-York), 1997.

\bibitem{F+07} J. Fuchs, M.R. Gaberdiel, I. Runkel, C. Schweigert, ``Topological defects for the free boson CFT", J. Phys. A: Math. Theor. 40 (2007) 11403.

\bibitem{GT14} K. Gawedzki, C. Tauber, ``Non-equilibrium transport in quantum wires through free massless bosonic fields", to appear.

\bibitem{Giama92} T. Giamarchi, A.J. Millis, ``Conductivity of a Luttinger liquid", Phys. Rev. {\bf B46} (1922) 9325.

\bibitem{GLSS06} S. Gustavsson, et al, ``Counting statistics of single electron transport in quantum dot", Phys. Rev. Lett. 96, 076605 (2006).

\bibitem{GGM10} D.B. Gutman, Yu. Gefen, A.D. Mirlin, ``Full counting statistics of Luttinger liquid conductor", Phys. Rev. Lett. 105, 256802 (2010).

\bibitem{Jak1} V. Jak\v{s}i\'c and C.-A. Pillet, ``On entropy production in quantum statistical mechanics'', Commun. Math. Phys. 217, 285Ð293
(2001)

\bibitem{Jak2} V. Jak\v{s}i\'c and C.-A. Pillet, ``A note on the entropy production formula'', Contemp. Math. 327, 175 (2003).

\bibitem{Pierre} S. Jezouin et al., ``Quantum Limit of Heat Flow Across a Single Electronic Channel", Science 342, 601 (2013).

\bibitem{KSH14} S. Krinner, D. Stadler, D. Husmann, J.-P. Brantut and T. Esslinger, \lq\lq Observation of Quantized Conductance in Neutral Matter\rq\rq,  ArXiv:1404.6400,
2014.

\bibitem{Moore12}  C. Karrasch R. Ilan and J. E. Moore, ``Non-equilibrium thermal transport and its relation to linear response", arXiv:1211.2236.

\bibitem{DVMR14} A. De Luca, J. Viti, L. Mazza and D. Rossini, ``Energy transport in Heisenberg chains beyond the Luttinger liquid paradigm'', Phys. Rev. B 90 16 (2014).
 
\bibitem{MCJ14} J. J. Mendoza-Arenas, S. R. Clark and D. Jaksch, ``Coexistence of energy diffusion and local thermalization in
nonequilibrium XXZ spin chains with integrability breaking'', ArXiv:1410.5838. 
 
\bibitem{MS13} M. Mintchev, P. Sorba, ``Luttinger Liquid in Non-equilibrium Steady State", J. Phys. A: Math. Theor. 46 (2013)

\bibitem{PZ01} V.B. Petkova, J.-B. Zuber, ``Generalised twisted partition functions", Phys. Lett. B 504 (2001) 533.

\bibitem{Pillet1} C.-A. Pillet, ``Entropy production in classical and quantum systems'', Markov Process. Relat. Fields 7, 145Ð157
(2001).

\bibitem{Pros11} T. Prosen, ``Open XXZ Spin Chain ``Non-equilibrium Steady State and a Strict Bound on Ballistic Transport", Phys. Rev. Lett. {\bf 106} (2011) 217206.

\bibitem{QRW07} T. Quella, I. Runkel, G.M.T. Watts, ``Reflection and transmission for conformal defects", JHEP04 (2007) 095.

\bibitem{Ruelle} D. Ruelle, ``Natural non-equilibrium states in quantum statistical mechanics", J. Stat. Phys. 98, 57 (2000).

\bibitem{Ruelle2} D. Ruelle, ``Entropy production in quantum spin systems'', Commun. Math. Phys. 224, 3Ð16 (2001).

\bibitem{SenFerm} D. Senechal, ``An introduction to bosonization", arXiv:cond-mat/9908262.

\bibitem{Aff05} J. Sirker, R.G. Pereira, I. Affleck, ``Diffusion and Ballistic Transport in One Dimensional Quantum Systems", Phys. Rev. Lett. {\bf 103} (2009) 216602

\bibitem{Zwanzig} R. Zwanzig,``Non-equilibrium statistical physics", Oxford Univ. Press 2002.

\end{thebibliography}
\end{document}